\documentclass[10pt,journal]{IEEEtran}

\usepackage{bm}
\usepackage{amssymb}
\usepackage{mathtools}
\usepackage{amsmath}
\usepackage{graphicx}
\usepackage{booktabs}
\usepackage{algorithm}
\usepackage{algpseudocode}
\usepackage[colorlinks=true, allcolors=blue]{hyperref}
\usepackage{enumitem}
\usepackage{tikz}
\usetikzlibrary{positioning, arrows.meta, decorations.pathreplacing, calc,
    patterns}

\algrenewcommand{\algorithmicrequire}{\textbf{Input:}}
\algrenewcommand{\algorithmicensure}{\textbf{Output:}}

\newtheorem{lemma}{Lemma}
\newtheorem{theorem}{Theorem}
\newtheorem{proposition}{Proposition}

\begin{document}

\title{From Partitioning to Provisioning: Co-Designing DNN Deployment and Infrastructure Capacity}
\author{Zongyao Huang, Fan Yang, Qi Cao, Yulin Shao
\thanks{Z. Huang and Q. Cao are with Guangzhou Institute of Technology, Xidian University, Guangzhou, China.}
\thanks{F. Yang and Y. Shao are with the Department of Electrical and Computer Engineering, The University of Hong Kong, Hong Kong, China.}
}

\maketitle

\begin{abstract}
Deep neural network (DNN) partitioning enables distributed inference by dividing model computation across interconnected devices. Existing approaches have largely focused on optimizing DNN deployment over infrastructure with fixed computing and communication capacities. Yet when these resources cannot adequately support inference demands, optimizing deployment alone is insufficient: the supporting infrastructure must also be provisioned. Crucially, deployment and provisioning are inherently coupled. How a DNN is partitioned determines where capacity is needed, while capacity provisioning can fundamentally change the preferred deployment. Motivated by this interdependence, we investigate the co-design of DNN deployment and infrastructure capacity for pipelined inference. Despite the coupling between these decisions, we reveal an efficiently solvable structure: on acyclic networks, the problem admits an exact shortest-path reformulation, enabling globally optimal joint design in polynomial time. Numerical evaluations demonstrate the benefits of co-design over treating deployment and provisioning separately, highlighting the potential of inference-driven infrastructure planning.
\end{abstract}

\begin{IEEEkeywords}
Distributed DNN inference, DNN partitioning, infrastructure co-design, capacity provisioning.
\end{IEEEkeywords}

\section{Introduction}
\label{sec:introduction}

Distributed deep neural network (DNN) inference~\cite{teerapittayanon2017distributed,zhao2018deepthings} enables computationally demanding models to be executed across interconnected devices. A key enabling technique is DNN partitioning~\cite{kang2017neurosurgeon,eshratifar2021jointdnn}, which distributes different portions of a model across devices. For streaming workloads, these portions can operate as a pipeline, allowing successive inputs to be processed concurrently~\cite{hadidi2020collaborative,hu2022pipeedge}. How to partition and deploy DNNs across computing networks has thus become a fundamental problem in distributed inference.

Existing research has largely approached this problem from a deployment perspective: given an infrastructure with fixed computing and communication capacities, how should a DNN be partitioned and deployed to achieve the best inference performance? Extensive efforts have been devoted to optimizing inference latency~\cite{kang2017neurosurgeon,li2020edgeai,hu2019dads}, throughput~\cite{zhao2018deepthings,hu2022pipeedge}, and other performance metrics~\cite{zeng2021coedge,laskaridis2020spinn} under such resource constraints. While this perspective addresses how to make effective use of existing infrastructure, it leaves open a complementary question: \emph{how should the infrastructure itself be provisioned to support the inference workloads deployed on it?}

The answer depends on the deployment itself. Partitioning determines where computation occurs and how much data traverses each link \cite{kang2017neurosurgeon,hu2019dads}, thereby shaping where additional capacity is needed. Yet capacity expansion can change which deployment is preferable: a deployment that maximizes throughput with existing resources may be far more expensive to upgrade than an alternative. Deployment and infrastructure capacity must therefore be designed jointly, rather than optimized in sequence.

In this paper, we develop such a co-design framework for pipelined DNN inference. Given a computing network with heterogeneous capacity costs, we jointly optimize the physical deployment path, the partitioning of DNN computational blocks, and the additional computing and communication capacities needed to sustain a target inference throughput. Although this problem couples discrete deployment choices with continuous provisioning decisions, we show that it possesses an exact and efficiently solvable structure on acyclic networks. The key insight is that the bottleneck-stage characterization of pipeline throughput allows the capacity requirements of individual stages to be determined locally. These requirements induce additive provisioning costs along a deployment path, enabling an exact reduction of the joint design problem to a shortest-path problem on a state graph.

The main contributions of this paper are summarized below:
\begin{itemize}[leftmargin=0.4cm]
    \item We formulate an inference-infrastructure co-design problem that jointly determines DNN partitioning, deployment paths, and computing and communication capacity increments. Unlike fixed-capacity deployment optimization, our formulation treats infrastructure capacity as a design variable, allowing inference deployment and capacity provisioning to be optimized together.
    \item We establish an exact equivalence between the co-design problem on acyclic networks and a shortest-path problem on an appropriately constructed state graph. Based on this structural result, we develop a dynamic programming algorithm that obtains a globally optimal solution in \(O(|\mathcal V|N+|\mathcal E|N^2)\) time and \(O(|\mathcal V|N)\) space.
    \item We evaluate the proposed approach on $400$ instances with VGG-16 and ResNet-50 workloads. Compared with two heuristic provisioning strategies that do not jointly optimize deployment and capacity expansion, our approach achieves substantial cost reductions, with median cost ratios reaching $25$ and worst-case ratios exceeding $100$. These results highlight the importance of co-designing DNN deployment and its supporting infrastructure capacity.
\end{itemize}

\section{Problem Formulation}
\label{sec:system-model}

We first describe how DNN deployment and network capacities determine pipelined inference throughput, and then formulate the minimum-cost co-design problem under a target throughput requirement.

\subsection{DNN Deployment Model}
\label{subsec:network-model}

The distributed computing network is modeled as a directed graph \(G=(\mathcal V,\mathcal E)\), where \(\mathcal V\) and \(\mathcal E\) denote the sets of physical nodes and directed links, respectively. The network contains a source node \(s\in\mathcal{V}\) and a destination node \(t\in\mathcal{V}\), with \(s\neq t\), and at least one path from \(s\) to \(t\) is assumed to exist. Each node \(v\in\mathcal{V}\) has existing computing capacity \(C_v>0\), measured in floating-point operations per second. Each link \(e\in\mathcal{E}\) has existing communication capacity \(B_e>0\), measured in bits per second. We collect the existing capacities into the vectors \(\bm C=(C_v)_{v\in\mathcal V}\) and \(\bm B=(B_e)_{e\in\mathcal E}\).

The DNN is represented by an ordered sequence of \(N\geq 1\) heterogeneous blocks, each of which may be a single layer or a module such as a residual or inception module. To characterize the computation and data transfer requirements, let \(Q_n>0\) denote the number of floating-point operations required by block \(n\), for \(n=1,\ldots,N\); and let \(D_n>0\) denote the activation size in bits after the first \(n\) blocks, for \(n=0,\ldots,N\). Thus, \(D_0\) is the raw input size and \(D_N\) is the final output size. To express the computing workload of consecutive blocks assigned to a node, we define the prefix workload \(S_0=0\) and \(S_k=\sum_{n=1}^{k}Q_n\) for \(k=1,\ldots,N\). A node executing blocks \(i+1,\ldots,j\) then requires \(S_j-S_i\) floating-point operations.

A DNN deployment specifies both the physical route through the network and how consecutive DNN blocks are assigned to nodes on that route. We represent it by a path \(p\) and a vector of partition points \(\bm b_p\). The path \(p\) is defined as
\begin{equation}
\label{eq:path-def}
\begin{aligned}
p&=(v_{p,1},v_{p,2},\ldots,v_{p,L_p}),\quad
v_{p,1}=s,\quad v_{p,L_p}=t,\\
&(v_{p,\ell},v_{p,\ell+1})\in\mathcal{E},\quad
\ell=1,\ldots,L_p-1,\\
&v_{p,i}\ne v_{p,j},\quad
\forall\,1\le i<j\le L_p,
\end{aligned}
\end{equation}
where \(L_p\) is the number of nodes on \(p\), and \(v_{p,\ell}\) denotes its \(\ell\)-th node. Let \(\mathcal{P}\) be the set of all such paths. For each \(p\in\mathcal{P}\), the vector of partition points is
\begin{equation}
\label{eq:bp-def}
\begin{aligned}
\bm b_p&=(b_{p,0},b_{p,1},\ldots,b_{p,L_p})\in\mathbb{N}_0^{L_p+1},\\
&0=b_{p,0}\leq b_{p,1}\leq\cdots\leq b_{p,L_p}=N.
\end{aligned}
\end{equation}
Let \(\mathcal B_p\) be the set of all such vectors, where \(\mathbb N_0\) denotes the non-negative integers. For \(\ell=1,\ldots,L_p\), \(b_{p,\ell}\) denotes the number of DNN blocks completed along \(p\) up to and including node \(v_{p,\ell}\). Accordingly, node \(v_{p,\ell}\) executes blocks \(b_{p,\ell-1}+1,\ldots,b_{p,\ell}\). If \(b_{p,\ell}=b_{p,\ell-1}\), the node performs no computation and only forwards the activation. After a node \(v_{p,\ell}\neq t\) completes its assigned blocks, it sends the activation of size \(D_{b_{p,\ell}}\) over the link \((v_{p,\ell},v_{p,\ell+1})\).

\subsection{Pipelined Throughput}
\label{subsec:throughput}

In pipelined execution, the inference consists of computation stages and communication stages. We consider steady-state pipelined inference with sufficient buffering and no resource contention, where computation and communication stages can overlap across successive inputs. The slowest stage determines the output interval, whose reciprocal gives the inference throughput~\cite{hu2019dads}.
For a deployment \((p,\bm b_p)\), the computation time at each node equals its assigned workload divided by its computing capacity. Hence, the maximum computation-stage delay is
\[
T^{\mathrm{comp}}(p,\bm b_p;\bm C)=\max_{\ell=1}^{L_p}\frac{S_{b_{p,\ell}}-S_{b_{p,\ell-1}}}{C_{v_{p,\ell}}}.
\]
For \(\ell=1,\ldots,L_p-1\), let \(e_{p,\ell}=(v_{p,\ell},v_{p,\ell+1})\) be the \(\ell\)-th link on \(p\). The corresponding communication time depends on the activation size and link capacity, giving the maximum communication-stage delay
\[
T^{\mathrm{comm}}(p,\bm b_p;\bm B)=\max_{\ell=1}^{L_p-1}\frac{D_{b_{p,\ell}}}{B_{e_{p,\ell}}}.
\]
The output interval is therefore
\[
T(p,\bm b_p;\bm C,\bm B)=\max\left\{T^{\mathrm{comp}}(p,\bm b_p;\bm C),T^{\mathrm{comm}}(p,\bm b_p;\bm B)\right\}.
\]
The minimum achievable output interval over all feasible deployments is
\begin{equation}
\label{eq:minimum-interval}
T^*(\bm C,\bm B)
=
\min_{p\in\mathcal P,\ \bm b_p\in\mathcal B_p}
T(p,\bm b_p;\bm C,\bm B),
\end{equation}
and the maximum achievable inference throughput under these capacities is \(1/T^*(\bm C,\bm B)\).

\subsection{The Minimum-Cost Co-Design Problem}
\label{subsec:provisioning-problem}

Suppose the inference service requires a throughput of at least \(1/\beta\), where \(\beta>0\), or equivalently, an output interval no greater than \(\beta\). The goal is to meet this requirement at minimum provisioning cost by jointly determining the DNN deployment and capacity increments.

For every node \(v\in\mathcal V\), let \(x_v\geq 0\) be the increment added to \(C_v\), and for every link \(e\in\mathcal E\), let \(y_e\geq 0\) be the increment added to \(B_e\). We collect these increments into the vectors \(\bm x=(x_v)_{v\in\mathcal V}\) and \(\bm y=(y_e)_{e\in\mathcal E}\). Since the cost of adding capacity can vary across the network, let \(\pi^{\mathrm{comp}}_v>0\) and \(\pi^{\mathrm{comm}}_e>0\) be the unit prices of computing capacity and communication capacity, respectively, and collect them into the vectors \(\bm\pi^{\mathrm{comp}}=(\pi^{\mathrm{comp}}_v)_{v\in\mathcal V}\) and \(\bm\pi^{\mathrm{comm}}=(\pi^{\mathrm{comm}}_e)_{e\in\mathcal E}\). We consider linear provisioning costs for simplicity. The proposed method also applies to costs that are separable across nodes and links, non-decreasing in each capacity increment, and zero at zero increment.

After provisioning, the required throughput is achievable if \(T^*(\bm C+\bm x,\bm B+\bm y)\leq\beta\). We therefore formulate the minimum-cost co-design problem as
\begin{equation}
\label{eq:provisioning}
\begin{aligned}
\min_{\bm x,\bm y}\quad & \left(\bm\pi^{\mathrm{comp}}\right)^{\mathsf T}\bm x+\left(\bm\pi^{\mathrm{comm}}\right)^{\mathsf T}\bm y\\
\text{s.t.}\quad & T^*(\bm C+\bm x,\bm B+\bm y)\leq\beta,\quad \bm x,\bm y\geq\bm 0.
\end{aligned}
\end{equation}

If the existing network already meets the throughput requirement, the optimal provisioning cost is zero. Otherwise, although \eqref{eq:provisioning} explicitly optimizes only \(\bm x\) and \(\bm y\), the physical path and partition points are implicitly optimized through \(T^*\) in \eqref{eq:minimum-interval}. Since the preferred deployment can change with variable capacities, deployment and provisioning decisions are coupled. In the next section, we exploit the acyclic network structure to obtain an exact solution to this co-design problem.

\section{Exact Co-Design on Acyclic Networks}
\label{sec:state-graph-reduction}

\subsection{Cost-Weighted State Representation}
\label{subsec:state-graph}

We represent the deployment decisions in \eqref{eq:provisioning} using an auxiliary state graph \(G_{\mathrm S}=(\mathcal V_{\mathrm S},\mathcal E_{\mathrm S})\), where \(\mathcal V_{\mathrm S}\) is the set of states and \(\mathcal E_{\mathrm S}\) is the set of state transitions, both defined later in this section. Two edge weight functions \(\delta(\cdot;\bm x,\bm y),\,w:\mathcal E_{\mathrm S}\to\mathbb R_{\geq0}\) are defined in Sections~\ref{subsec:delay} and~\ref{subsec:cost}, respectively.

Introduce a virtual terminal node \(\tau\notin\mathcal V\).
For each node, let \(\mathcal I_v\) denote its feasible block indices, with \(\mathcal I_s=\{0\}\), \(\mathcal I_v=\{0,1,\ldots,N\}\) for every \(v\in\mathcal V\setminus\{s\}\), and \(\mathcal I_\tau=\{N\}\). The set of states (vertices of \(G_\mathrm{S}\)) is therefore \(\mathcal V_{\mathrm S}=\{(v,i):v\in\mathcal V\cup\{\tau\},\ i\in\mathcal I_v\}\). A state \((v,i)\) means that the activation after block \(i\) has arrived at \(v\), before \(v\) processes any further blocks. In particular, \((s,0)\) means that the raw input has arrived at \(s\) and no DNN blocks have yet been executed; \((\tau,N)\) means that all \(N\) blocks have been processed and the final output is available at \(t\).

Introduce a virtual link \(e_\tau=(t,\tau)\) with \(B_{e_\tau}=+\infty\) and \(y_{e_\tau}=\pi^{\mathrm{comm}}_{e_\tau}=0\), so that the weight formulas below apply to \(e_\tau\). The set of state transitions (edges of \(G_\mathrm{S}\)) is
\begin{equation}
\label{eq:transition-set}
\begin{aligned}
\mathcal E_{\mathrm S}
=
\bigl\{((v,i),(z,j)):\;&(v,z)\in\mathcal E\cup\{e_\tau\},\\
&i\in\mathcal I_v,\ j\in\mathcal I_z,\ i\leq j
\bigr\}.
\end{aligned}  
\end{equation}
For \(\xi=\big((v,i),(z,j)\big)\in\mathcal E_{\mathrm S}\), define \(v(\xi)\triangleq v\), \(e(\xi)\triangleq (v,z)\), \(i(\xi)\triangleq i\), \(j(\xi)\triangleq j\).

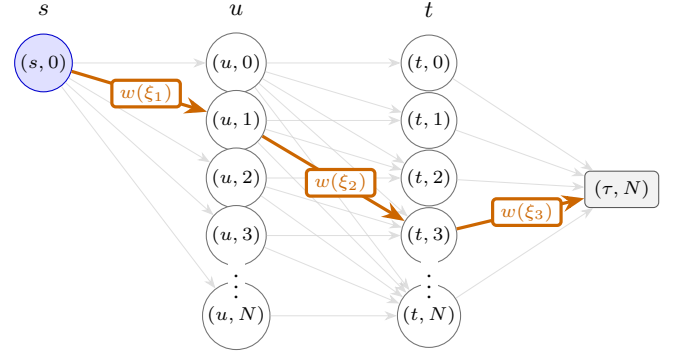
\begin{figure}[t]
    \centering
    \begin{tikzpicture}[
        >=Stealth,
        state/.style={
            circle, draw=black!55, fill=white, minimum size=0.62cm,
            inner sep=1pt, font=\scriptsize
        },
        source/.style={
            circle, draw=blue!80!black, fill=blue!12, minimum size=0.68cm,
            inner sep=1pt, font=\scriptsize
        },
        sink/.style={
            rectangle, rounded corners=2pt, draw=black!65,
            fill=gray!10, minimum width=0.72cm, minimum height=0.48cm,
            font=\scriptsize
        },
        lightarc/.style={
            ->, draw=gray!35, line width=0.38pt, opacity=0.72
        },
        selectedpath/.style={->, draw=orange!80!black, line width=1.2pt},
        rlabel/.style={
            rounded corners=1.5pt, draw=orange!80!black, fill=white,
            inner xsep=3pt, inner ysep=1.5pt,
            font=\scriptsize, text=orange!80!black
        },
        every node/.style={align=center}
    ]
        \node[font=\small\bfseries] at (0,0.70) {\(s\)};
        \node[source] (s0) at (0,0) {\((s,0)\)};
        \foreach \x/\name/\label in {2.55/u/u,5.10/t/t} {
            \node[font=\small\bfseries] at (\x,0.70) {\(\label\)};
            \foreach \i in {0,1,2,3} {
                \node[state] (\name\i) at (\x,{-0.76*\i}) {\((\label,\i)\)};
            }
            \node[state] (\name N) at (\x,-3.34) {\((\label,N)\)};
        }
        \node[sink] (tau) at (7.65,-1.67) {\((\tau,N)\)};

        \foreach \j in {0,1,2,3} {
            \draw[lightarc] (s0) -- (u\j);
        }
        \draw[lightarc] (s0) -- (uN);
        \foreach \i in {0,1,2,3} {
            \foreach \j in {\i,...,3} {
                \draw[lightarc] (u\i) -- (t\j);
            }
            \draw[lightarc] (u\i) -- (tN);
        }
        \draw[lightarc] (uN) -- (tN);
        \foreach \i in {0,1,2,3} {
            \draw[lightarc] (t\i) -- (tau);
        }
        \draw[lightarc] (tN) -- (tau);
        \foreach \x in {2.55,5.10} {
            \node[font=\small, fill=white, inner sep=2pt] at (\x,-2.85)
                {\(\vdots\)};
        }
        \draw[selectedpath] (s0) -- node[rlabel, pos=0.55] {\(w(\xi_1)\)} (u1);
        \draw[selectedpath] (u1) -- node[rlabel, pos=0.55] {\(w(\xi_2)\)} (t3);
        \draw[selectedpath] (t3) -- node[rlabel, pos=0.55] {\(w(\xi_3)\)} (tau);
    \end{tikzpicture}
    \caption{The state graph \(G_{\mathrm S}\) for a small example with one intermediate node \(u\) and links \((s,u)\), \((u,t)\). Each transition carries the cost weight \(w(\xi)\) needed to meet the throughput constraint. The highlighted state path selects the physical path and the partition points. The vertical dots represent omitted states with indices \(4,\ldots,N-1\) and their incident transitions.}
    \label{fig:state-graph}
\end{figure}

We call a simple path from \((s,0)\) to \((\tau,N)\) in \(G_{\mathrm S}\) a state path, and let \(\mathcal R\) be the set of all state paths. For each \(r\in\mathcal R\), let \(\mathcal E_{\mathrm S}(r)\subseteq\mathcal E_{\mathrm S}\) be the set of transitions between states of \(r\).
Fig.~\ref{fig:state-graph} illustrates \(G_{\mathrm S}\) for a simple network. We next define the delay and provisioning cost associated with each state transition.

\subsubsection{Delay of Transitions}
\label{subsec:delay}

For a transition \(\xi\in\mathcal E_{\mathrm S}\), the computation at \(v(\xi)\) and communication over \(e(\xi)\) form two pipeline stages. Under capacity increments \(\bm x\) and \(\bm y\), its delay is represented as the larger of their delays:
\[
\delta(\xi;\bm x,\bm y)
\triangleq
\max\left\{
\frac{S_{j(\xi)}-S_{i(\xi)}}{C_{v(\xi)}+x_{v(\xi)}},
\frac{D_{j(\xi)}}{B_{e(\xi)}+y_{e(\xi)}}
\right\}.
\]

The following lemma shows that state paths preserve both the deployment choices and their output intervals.

\begin{lemma}
\label{lem:correspondence}
If \(G\) is acyclic, the mapping \((p,\bm b_p)\mapsto r\) defines a bijection between
\(
\{(p,\bm b_p):p\in\mathcal P,\ \bm b_p\in\mathcal B_p\}
\)
and \(\mathcal R\). Moreover, if \(r\) is the state path corresponding to \((p,\bm b_p)\), then \(T(p,\bm b_p;\bm C+\bm x,\bm B+\bm y)=\max_{\xi\in\mathcal E_{\mathrm S}(r)}\delta(\xi;\bm x,\bm y)\).
\end{lemma}

\begin{IEEEproof}
\emph{Forward map.} Given \((p,\bm b_p)\), let \(v_{p,L_p+1}=\tau\) and define \(\eta_\ell=(v_{p,\ell},b_{p,\ell-1})\) for \(\ell=1,\ldots,L_p+1\). By \eqref{eq:path-def} and \eqref{eq:bp-def}, each consecutive pair \((\eta_\ell,\eta_{\ell+1})\) belongs to \(\mathcal E_{\mathrm S}\). The resulting sequence starts at \((s,0)\) and ends at \((\tau,N)\), and thus defines a state path \(r\in\mathcal R\).

\emph{Inverse map.} Let \(r\in\mathcal R\) be represented by the sequence \(\big((u_1,i_1),\dots,(u_{k-1},i_{k-1}),(u_k,i_k)\big)\), where \(u_1=s\), \(i_1=0\), \(u_{k-1}=t\), \(u_k=\tau\) and \(i_k=N\). By \eqref{eq:transition-set}, \((u_\ell,u_{\ell+1})\in\mathcal E\cup\{e_\tau\}\) for all \(\ell=1,\ldots,k-1\). Since \(G\) is acyclic, \(u_1,\dots,u_{k-1}\) are pairwise distinct, and hence \(p=(u_1,\dots,u_{k-1})\in\mathcal P\). Moreover, since \(i_\ell\le i_{\ell+1}\) for all \(\ell=1,\ldots,k-1\), \(\bm b_p=(i_1,\dots,i_{k-1},i_k)\in\mathcal B_p\). Thus, the forward and inverse constructions are mutually inverse, proving the bijection.

\emph{Interval equality.}  Each transition \(\xi_\ell=(\eta_\ell,\eta_{\ell+1})\) represents the computation at node \(v_{p,\ell}\) followed by transmission over its outgoing link. Let \(e_{p,L_p}=e_\tau\). Since \(B_{e_\tau}=+\infty\), the final transition has zero communication delay. Therefore,
\[
\begin{aligned}
&T(p,\bm b_p;\bm C+\bm x,\bm B+\bm y)\\
&=\max_{\ell=1}^{L_p}
\max\left\{
\frac{S_{b_{p,\ell}}-S_{b_{p,\ell-1}}}
{C_{v_{p,\ell}}+x_{v_{p,\ell}}},
\frac{D_{b_{p,\ell}}}
{B_{e_{p,\ell}}+y_{e_{p,\ell}}}
\right\}\\
&=\max_{\xi\in\mathcal E_{\mathrm S}(r)}\max\left\{
\frac{S_{j(\xi)}-S_{i(\xi)}}{C_{v(\xi)}+x_{v(\xi)}},
\frac{D_{j(\xi)}}{B_{e(\xi)}+y_{e(\xi)}}
\right\}\\
&=\max_{\xi\in\mathcal E_{\mathrm S}(r)}
\delta(\xi;\bm x,\bm y).
\end{aligned}
\]
\end{IEEEproof}

By \eqref{eq:minimum-interval} and Lemma~\ref{lem:correspondence}, the minimum output interval can be expressed entirely over state paths:
\begin{equation}
\label{eq:bottleneck}
T^*(\bm C+\bm x,\bm B+\bm y)
=
\min_{r\in\mathcal R}
\max_{\xi\in\mathcal E_{\mathrm S}(r)}
\delta(\xi;\bm x,\bm y).
\end{equation}

\subsubsection{Cost of Transitions}
\label{subsec:cost}

By \eqref{eq:bottleneck}, a state path meets the throughput target if every transition on it has a delay no greater than \(\beta\). For each \(\xi\in\mathcal E_{\mathrm S}\), \(\delta(\xi;\bm x,\bm y)\leq\beta\) is equivalent to
\[
x_{v(\xi)}\geq\frac{S_{j(\xi)}-S_{i(\xi)}}{\beta}-C_{v(\xi)},\quad
y_{e(\xi)}\geq\frac{D_{j(\xi)}}{\beta}-B_{e(\xi)}.
\]
Together with the non-negativity constraints, these inequalities give the minimum computing and communication capacity increments required by \(\xi\):
\[
\begin{aligned}
\rho(\xi)&=\left[\frac{S_{j(\xi)}-S_{i(\xi)}}{\beta}-C_{v(\xi)}\right]^+,\\
\sigma(\xi)&=\left[\frac{D_{j(\xi)}}{\beta}-B_{e(\xi)}\right]^+,
\end{aligned}
\]
where \([x]^+\triangleq\max\{0,x\}\). Using the unit capacity prices, we define the provisioning cost of transition \(\xi\) as
\[
w(\xi)
\triangleq
\pi^{\mathrm{comp}}_{v(\xi)}\,\rho(\xi)
+
\pi^{\mathrm{comm}}_{e(\xi)}\,\sigma(\xi).
\]

\subsection{An Exact Shortest-Path Solution}
\label{subsec:reduction}

Lemma~\ref{lem:correspondence} establishes a correspondence between DNN deployments and state paths, while \(w(\xi)\) gives the minimum provisioning cost for each transition. If the physical network is acyclic, no physical node or link is used more than once along a state path. These costs can therefore be added along the path, leading to the following exact equivalence.

\begin{theorem}
\label{thm:provisioning}
If \(G\) is acyclic, the optimal value of problem~\eqref{eq:provisioning} equals the shortest path distance from \((s,0)\) to \((\tau,N)\) in \(G_{\mathrm S}\) with \(w\) as the edge weight function, i.e., \(\min_{r\in\mathcal R}\sum_{\xi\in\mathcal E_{\mathrm S}(r)}w(\xi)\). Moreover, if a shortest state path is found, an optimal solution of~\eqref{eq:provisioning} can be derived accordingly.
\end{theorem}

\begin{IEEEproof}
For any transition \(\xi\in\mathcal E_{\mathrm S}\), the delay requirement and non-negative capacity increments give
\begin{equation}
\label{eq:threshold}
\delta(\xi;\bm x,\bm y)\leq\beta
\iff
x_{v(\xi)}\geq\rho(\xi),\quad
y_{e(\xi)}\geq\sigma(\xi).
\end{equation}
Let \(J(\bm x,\bm y)=(\bm\pi^{\mathrm{comp}})^{\mathsf T}\bm x+(\bm\pi^{\mathrm{comm}})^{\mathsf T}\bm y\). For any feasible \((\bm x,\bm y)\), \eqref{eq:bottleneck} guarantees a state path \(r\in\mathcal R\) satisfying \(\delta(\xi;\bm x,\bm y)\leq\beta\) for every \(\xi\in\mathcal E_{\mathrm S}(r)\). Since \(G\) is acyclic, no physical node or link appears more than once on \(r\). Thus, by \eqref{eq:threshold} and the non-negative capacity prices,
\[
J(\bm x,\bm y)
\geq \sum_{\xi\in\mathcal E_{\mathrm S}(r)}w(\xi)
\geq \min_{r'\in\mathcal R}
\sum_{\xi\in\mathcal E_{\mathrm S}(r')}w(\xi).
\]
Conversely, let \(r^*\) be a shortest state path, and set \(x^*_{v(\xi)}=\rho(\xi)\) and \(y^*_{e(\xi)}=\sigma(\xi)\) for all \(\xi\) of \(r^*\), with all other increments set to zero. By \eqref{eq:threshold}, this allocation is feasible. Since no physical resource is repeated along \(r^*\),
\[
J(\bm x^*,\bm y^*)
=\sum_{\xi\in\mathcal E_{\mathrm S}(r^*)}w(\xi)
=\min_{r\in\mathcal R}
\sum_{\xi\in\mathcal E_{\mathrm S}(r)}w(\xi).
\]
The corresponding deployment follows from Lemma~\ref{lem:correspondence}.
\end{IEEEproof}

\begin{algorithm}[t]
\caption{Provisioning on a State Graph (PSG)}
\label{alg:provisioning}
\begin{algorithmic}[1]
\small
\Require \(G=(\mathcal V,\mathcal E)\), \(\bm C\), \(\bm B\), \(Q_1,\ldots,Q_N\), \(D_0,\ldots,D_N\), \(\beta\), \(\bm\pi^{\mathrm{comp}}\), \(\bm\pi^{\mathrm{comm}}\)
\Ensure optimal physical path \(p^*\), partition points \(\bm b_{p^*}\), and capacity increments \(\bm x^*\), \(\bm y^*\)
\State compute \(S_0,\ldots,S_N\) and a topological order \(v_1,\ldots,v_{|\mathcal V|}\) of \(\mathcal V\); set \(v_{|\mathcal V|+1}\leftarrow\tau\)
\State \(\mathrm{dist}(s,0)\leftarrow0\), \(\mathrm{dist}(v,i)\leftarrow+\infty\) for all other states, and clear \(\mathrm{pred}\)
\For{\(d=1,\ldots,|\mathcal V|+1\)}
\ForAll{\(i\in\mathcal I_{v_d}\) with \(\mathrm{dist}(v_d,i)<+\infty\)}
\ForAll{\((v_d,z)\in\mathcal E\cup\{e_\tau\}\)}
\ForAll{\(j\in\mathcal I_z\) with \(j\geq i\)}
\State \(\xi\leftarrow\bigl((v_d,i),(z,j)\bigr)\)
\State compute \(w(\xi)\)
\If{\(\mathrm{dist}(v_d,i)+w(\xi)<\mathrm{dist}(z,j)\)}
\State \(\mathrm{dist}(z,j)\leftarrow\mathrm{dist}(v_d,i)+w(\xi)\)
\State \(\mathrm{pred}(z,j)\leftarrow(v_d,i)\)
\EndIf
\EndFor
\EndFor
\EndFor
\EndFor
\State backtrack \(\mathrm{pred}\) from \((\tau,N)\) to \((s,0)\) into a state path \(r^*\)
\State \Return \((p^*,\bm b_{p^*},\bm x^*,\bm y^*)\) from \(r^*\) as in the proofs of Lemma~\ref{lem:correspondence} and Theorem~\ref{thm:provisioning}, of cost \(\mathrm{dist}(\tau,N)\)
\end{algorithmic}
\end{algorithm}

By Theorem~\ref{thm:provisioning}, the co-design problem reduces to a shortest-path problem on \(G_{\mathrm S}\) when \(G\) is acyclic. Since every state transition follows a physical link or \(e_\tau\), a topological order of \(G\), with \(\tau\) placed last, also gives a topological order of the state graph. We can therefore solve the problem by dynamic programming without explicitly constructing \(G_{\mathrm S}\). Maintain a label \(\mathrm{dist}(v,i)\), set to \(0\) at \((s,0)\) and \(+\infty\) elsewhere. Each transition \(\xi=((v,i),(z,j))\) is relaxed according to
\begin{equation}
\label{eq:relaxation}
\mathrm{dist}(z,j)\leftarrow
\min\bigl\{
\mathrm{dist}(z,j),\
\mathrm{dist}(v,i)+w(\xi)
\bigr\}.
\end{equation}
Processing states in topological order yields the optimal cost \(\mathrm{dist}(\tau,N)\). The predecessor pointers recover the corresponding deployment and capacity increments, as detailed in Algorithm~\ref{alg:provisioning}.

\begin{proposition}
\label{prop:complexity}
Algorithm~\ref{alg:provisioning} returns an optimal solution of \eqref{eq:provisioning} in \(O(|\mathcal V|N+|\mathcal E|N^2)\) time and \(O(|\mathcal V|N)\) space.
\end{proposition}

\begin{IEEEproof}
Optimality follows from Theorem~\ref{thm:provisioning}, since the relaxations in \eqref{eq:relaxation} are performed in topological order. The state graph \(G_{\mathrm S}\) has \(O(|\mathcal V|N)\) states and \(O(|\mathcal E|N^2)\) transitions. Each transition is processed once, and its weight is computed in \(O(1)\) time. Since transitions are generated on demand from \(\mathcal E\), the state graph need not be explicitly stored. Thus, only \(O(|\mathcal V|N)\) distance labels and predecessor pointers are required.
\end{IEEEproof}

The same algorithm applies to the separable, non-decreasing provisioning costs described in Section~\ref{subsec:provisioning-problem}. It suffices to redefine \(w(\xi)=f_{v(\xi)}(\rho(\xi))+g_{e(\xi)}(\sigma(\xi))\), where \(f_v\) and \(g_e\) are the corresponding cost functions, both non-decreasing on \(\mathbb R_{\geq0}\) and zero at zero.

\section{Numerical Results}
\label{sec:numerical-results}

We evaluate the benefit of minimum-cost co-design by comparing PSG (Algorithm~\ref{alg:provisioning}) with two provisioning heuristics that use throughput-optimimal deployments to guide capacity upgrades. The experiments consider randomly generated tiered networks with VGG-16 and ResNet-50 workloads.

\begin{algorithm}[t]
\caption{Provisioning with Water-Filling (PWF)}
\label{alg:pwf}
\begin{algorithmic}[1]
\small
\Require \(G\), \(\bm C\), \(\bm B\), \(Q_1,\ldots,Q_N\), \(D_0,\ldots,D_N\), \(\beta\)
\Ensure \(\bm x\), \(\bm y\) with \(T^*(\bm C+\bm x,\bm B+\bm y)\leq\beta\)
\State \(\bm x,\bm y\leftarrow\bm 0\); solve \eqref{eq:minimum-interval} for \(T^*\) and \((p,\bm b_p)\)
\While{\(T^*>\beta\)}
\State \(d\leftarrow\max\bigl(\{\beta\}\cup\{\lambda<T^*:\lambda\text{ is a stage delay of }(p,\bm b_p)\}\bigr)\)
\State let \(r\) be the state path corresponding to \((p,\bm b_p)\)
\ForAll{\(\xi\in\mathcal E_{\mathrm S}(r)\)}
\State \(x_{v(\xi)}\leftarrow\max\{x_{v(\xi)},\rho_d(\xi)\}\)
\If{\(e(\xi)\in\mathcal E\)}
\State \(y_{e(\xi)}\leftarrow\max\{y_{e(\xi)},\sigma_d(\xi)\}\)
\EndIf 
\EndFor
\State recompute \(T^*\) and \((p,\bm b_p)\) under \((\bm x,\bm y)\)
\EndWhile
\State \Return \(\bm x,\bm y\)
\end{algorithmic}
\end{algorithm}

Both heuristics use throughput-optimal deployments to guide provisioning, without accounting for capacity prices. \emph{PFP} (Provisioning under Frozen-Partitioning) fixes an initial throughput-optimal deployment and purchases the minimum capacity increments needed to meet \(\beta\). \emph{PWF} (Provisioning with Water-Filling) reduces the bottleneck stage delays of the current deployment to the next lower delay level or \(\beta\), and recomputes a throughput-optimal deployment after each upgrade. When \eqref{eq:minimum-interval} admits multiple throughput-optimal deployments, PWF selects one of them (the first found by the solver) rather than enumerating all optima. For any \(d>0\), let \(\rho_d(\xi)\) and \(\sigma_d(\xi)\) denote \(\rho(\xi)\) and \(\sigma(\xi)\) in Section~\ref{subsec:cost} with \(\beta\) replaced by \(d\). Algorithm~\ref{alg:pwf} summarizes PWF, which terminated on all tested instances.

The DNN workloads are analytic block profiles of VGG-16 (\(N=16\)) and ResNet-50 (\(N=18\)) with \(224\times224\) inputs. Block workloads and activation sizes are derived from layer shapes, assuming FP32 activations and two floating-point operations per multiply-accumulate, with max-pooling folded into
the preceding VGG-16 block and global average pooling folded into the last ResNet-50 block.

We generate 50 tiered DAGs, each consisting of a source \(s\), three intermediate tiers (access, middle, and remote) with 2--3 nodes each, and a destination \(t\). Each node has 1--3 randomly selected links to the next tier, except remote-tier nodes, which each connect to \(t\). The resulting networks contain 8--11 nodes and at least one path from \(s\) to \(t\). Existing capacities and unit prices follow the distributions in Table~\ref{tab:params}.

\begin{table}[t]
\centering
\caption{Capacities and unit prices (\(U\): uniform; \(\mathrm{LU}\): log-uniform).}
\label{tab:params}
\footnotesize
\begin{tabular}{@{}lll@{}}
\toprule
Entity & Distribution & Unit \\
\midrule
\(C_v\), source \& access tier & \(U[6,30]\) & GFLOP/s \\
\(C_v\), middle tier & \(U[30,90]\) & GFLOP/s \\
\(C_v\), remote tier \& \(t\) & \(U[90,180]\) & GFLOP/s \\
\(B_e\), links to access tier & \(U[30,100]\) & Mbps \\
\(B_e\), links to middle tier & \(U[100,300]\) & Mbps \\
\(B_e\), links to remote tier \& \(t\) & \(U[300,1000]\) & Mbps \\
\(\pi^{\mathrm{comp}}_v\), nodes & \(\mathrm{LU}[0.1,10]\) & per GFLOP/s \\
\(\pi^{\mathrm{comm}}_e\), links & \(\mathrm{LU}[0.01,1]\) & per Mbps \\
\bottomrule
\end{tabular}
\end{table}

For each topology and DNN, we set \(\beta=\alpha T^*(\bm C,\bm B)\), where \(\alpha\in\{0.25,0.4,0.6,0.8\}\). Since \(\alpha<1\), all instances require additional capacity. Using the same 50 topologies for both DNNs yields \(50\times2\times4=400\) instances. We measure the ratio of each heuristic's provisioning cost to the optimal cost obtained by PSG.

\begin{figure}[t]
\centering
\includegraphics[width=0.8\linewidth]{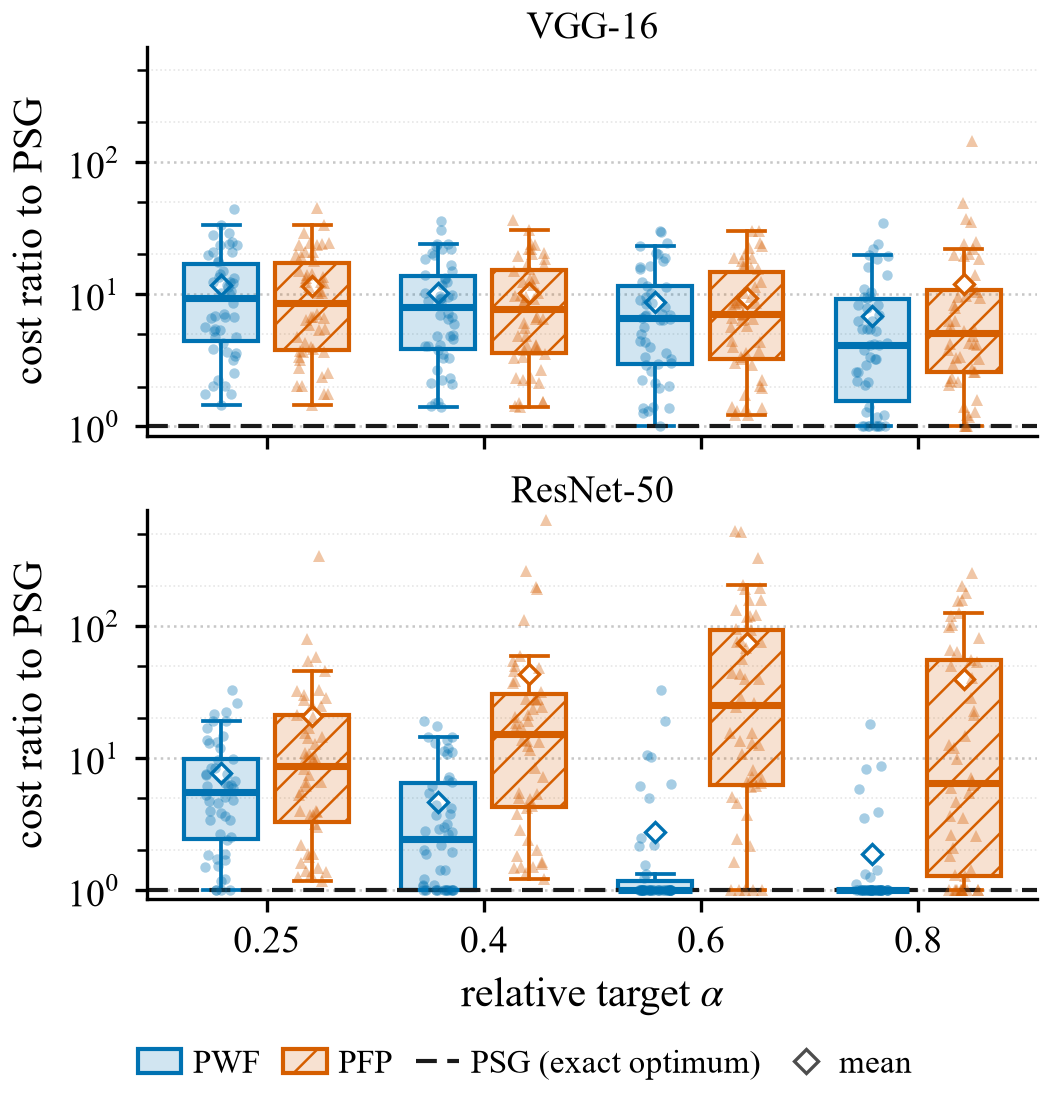}
\caption{The cost ratio of PWF and PFP to PSG versus the relative target \(\alpha\) on all 400 instances: VGG-16 (top) and ResNet-50 (bottom). Boxes span interquartile ranges, whiskers extend to \(1.5\times\) the interquartile range, diamonds mark means, dots mark individual instances, and the dashed line marks the exact optimum; the vertical axis is logarithmic.}
\label{fig:ensemble}
\end{figure}

Fig.~\ref{fig:ensemble} compares the provisioning-cost ratios over all 400 instances. PFP incurs median ratios of 5.1--8.6 for VGG-16 and 6.5--25.3 for ResNet-50, with worst cases of 144 and 639, respectively. PWF matches the optimum in 80\% of ResNet-50 instances at \(\alpha=0.8\), but its median ratios remain 4.1--9.4 for VGG-16 and reach 5.6 for ResNet-50, with a worst case of 44. For VGG-16, the median ratios of both heuristics increase as \(\alpha\) decreases, reaching 8.6 (PFP) and 9.4 (PWF) at \(\alpha=0.25\).

The cost gap reflects a key difference: the heuristics choose capacity upgrades without considering prices, whereas PSG minimizes price-weighted costs across feasible deployments. These results show that joint optimization can lead to much lower provisioning costs, even when deployment is updated during provisioning. In our Python implementation on an AMD Ryzen 7~8845H CPU, the median runtimes are 0.25, 0.28, and 1.8~ms for PSG, PFP, and PWF, respectively, showing that PSG remains computationally efficient on the tested instances.

\section{Conclusion}
\label{sec:conclusion}
This paper established an exact approach to co-designing DNN deployment and infrastructure capacity over acyclic networks. Beyond the algorithmic result, the broader insight is that the value of computing and communication capacity depends on how it serves the inference workload as a whole, rather than on individual resources in isolation. This points toward a fundamental shift in AI infrastructure planning: from expanding available resources to strategically provisioning capacity where it delivers the greatest end-to-end benefit.

\bibliographystyle{IEEEtran}
\bibliography{ref}

\end{document}